\documentclass{moriond}

\def\be{\begin{equation}}
\def\ee{\end{equation}}
\def\bea{\begin{eqnarray}}
\def\eea{\end{eqnarray}}

\newcommand{\Photo}{\includegraphics[height=35mm]{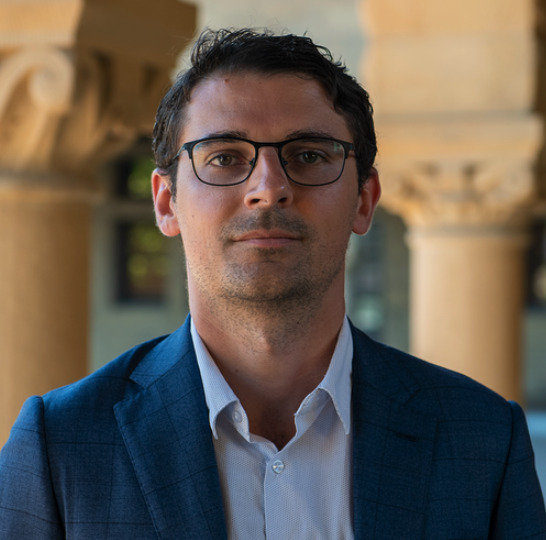}}

\begin{document}
\vspace*{4cm}
\title{Commissioning of the Vera C. Rubin Observatory}

\author{ Pierre-Fran\c{c}ois L\'eget}

\address{Department of Astrophysical Sciences, Princeton University,\\
Princeton, NJ 08544, USA}

\maketitle\abstracts{
The Vera C. Rubin Observatory began commissioning its camera, LSSTCam, in April 2025,
with the Legacy Survey of Space and Time (LSST) scheduled to start in 2026. A primary science goal is
constraining Dark Energy through weak gravitational lensing of the large-scale structure (cosmic shear). After a full year of
data from LSST, these measurements are expected to reach precision comparable to recent Dark Energy Spectroscopic Instrument (DESI), 
providing an independent test of hints that Dark Energy may evolve over time.
However, cosmic shear requires exquisite control of instrumental systematics. This
proceeding presents an overview of the Rubin Observatory commissioning — the successes achieved and
the systematic issues we are working to resolve.}

\section{The Vera C. Rubin Observatory}

The Vera C. Rubin Observatory is an 8.4-meter aperture optical wide-field survey
telescope located at Cerro Pach\'on (2,682~m) in Chile (Ivezi\'c et al. 2019~\cite{Ivezic2019}). Commissioning began in late
2024 with a commissioning camera, LSSTComCam (0.5 $\mathrm{deg}^2$), whose data were released
as Data Preview~1 (DP1) in mid-2025 (the Vera C. Rubin Observatory Team 2026~\cite{DP1Paper}). The main camera,
LSSTCam (9.6 $\mathrm{deg}^2$), started commissioning in April 2025 and has been operating ever since. 
The first batch of LSSTCam data is expected to be published as Data Preview~2 (DP2) (Guy et al. 2025~\cite{Guy2025})
in mid-2026, with real-time alerts having begun in February 2026. The Legacy Survey
of Space and Time (LSST) is scheduled to start in 2026.

The Rubin Observatory has broad scientific objectives spanning from solar system
physics to constraining the properties of Dark Energy. On this last topic, the
Rubin Observatory is expected to constrain dark energy properties through multiple cosmological probes. Among them, the $3\times2$pt analysis, combining cosmic shear, galaxy-galaxy lensing,
and galaxy clustering, is expected to reach the best precision alone. It should reach a precision
comparable to the current combination of DESI and CMB results after just one
year of the LSST survey --- provided systematic effects are under control.
However, early LSSTCam data show this is more challenging than anticipated.

In this proceeding, we discuss how image quality and Point Spread Function (PSF) modeling challenge cosmic shear
measurements, what we discovered once on sky, how we have improved, and
what remains to be done. Despite these challenges, we show that we can already
detect a weak lensing signal with no obvious PSF contamination in the case of
galaxy cluster lensing, which has less strict requirements than cosmic shear.

\section{On PSF modeling and image quality}

The modeling of the PSF is one of many systematic effects that can affect cosmic shear measurements
(see, e.g., Mandelbaum 2018~\cite{Mandelbaum2018}). Recent Dark
Energy constraints have included dedicated analyses to study the impact of PSF
systematics on $3\times2$pt measurements (Schutt et al. 2025~\cite{Schutt2025}, Yamamoto et al. 2025~\cite{Yamamoto2025}).
The general principle is to achieve an accurate description of second and higher
order moments (Zhang et al. 2023~\cite{Zhang2023}, Schutt et al. 2025~\cite{Schutt2025}) and to minimize their
correlations in sky coordinates (Schutt et al. 2025~\cite{Schutt2025}). Even when the PSF model
is well described, image quality remains an important metric. Visits with poor
image quality decrease survey depth, thereby raising the detection threshold for
faint sources such as high-redshift galaxies, decrease the number of resolved galaxies, and overall reduce the statistical power
of the survey (Dark Energy Science Collaboration (DESC) SRD 2018~\cite{DESCSRD2018}).

The Rubin Observatory PSF model is built within the LSST Science Pipelines (The Rubin Observatory Science Pipelines Team 2025~\cite{SciencePipelines2025}) using PSFs In the Full Field of view (PIFF; Jarvis et al. 2021~\cite{Jarvis2021}) and
is based on the same framework used for the Dark Energy Survey (DES) Y6 PSF model. During commissioning,
the main limitation to image quality was optical aberrations, which are controlled
by the Active Optics System (AOS) described in Megias Homar et al. 2024~\cite{MegiasHomar2024}. Both PSF
modeling and image quality during the commissioning phase did not perform as
expected for the LSST survey. The following sections present a walkthrough of the
problems encountered, how we addressed them, what remains challenging, and upcoming
improvements to PSF modeling and image quality.

\section{On improving PSF and image quality}

The data presented here are LSSTCam observations from the period that will form the basis of Data Preview~2, which was not yet public at the
time of writing but will be released in mid-2026. It corresponds to images acquired
between April 2025 and January 2026, totaling approximately 30,000 single-visit
exposures across the six filters of the Rubin Observatory. These data are used
throughout this proceeding to evaluate both PSF modeling and, indirectly, image
quality through PSF residuals.

\subsection{PSF residuals in focal plane coordinates}

A standard way to quantify biases in PSF modeling is by examining averages of PSF
second moment residuals in focal plane coordinates, as done for example in DES
Y6 (Schutt et al. 2025~\cite{Schutt2025}). Here we focus on residuals of the trace of the second moment
matrix $T$, defined as:
\be
\frac{\delta T}{T} = \frac{T_{\rm star} - T_{\rm PSF}}{T_{\rm star}}
\ee
where $T_{\rm star}$ is measured on stars used for training or validating the PSF
model, and $T_{\rm PSF}$ is estimated directly from the PSF model. DESC has strict
requirements on $\delta T/T$: below 0.4\% for LSST Year~1 and below 0.1\% for LSST
Year~10 (DESC SRD 2018~\cite{DESCSRD2018}). In the following, we focus mostly on $\delta T/T$, but ellipticity residuals
exhibit similar patterns. The analysis presented in this section and the next is
based on $\sim$3,000 visits, which was approximately what was available after full
pipeline processing in the pre-DP2 era.

\begin{figure}
    \centering
    \includegraphics[scale=0.25]{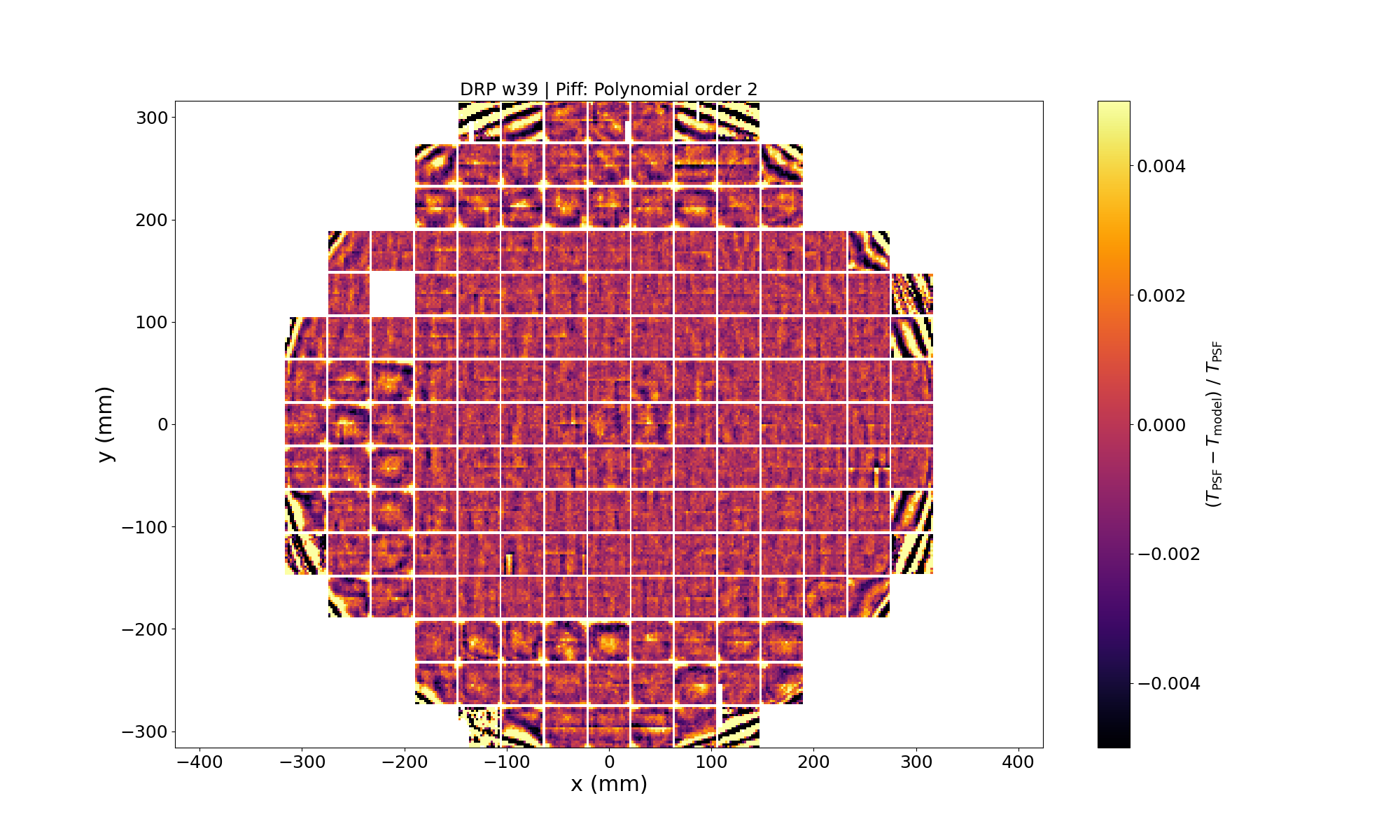}
    \vspace*{-4mm}
    \caption{$\delta T/T$ projected in focal plane coordinates for a subset of Data Preview~2. No image quality cut is applied. The PSF model uses a 2nd order polynomial interpolation.}
  \label{dTTFoV}
\end{figure}

Figure~\ref{dTTFoV} shows $\delta T/T$ averaged over these visits and projected into
focal plane coordinates. The focal plane is paved with two distinct types of Charge-Coupled Devices (CCDs)
from different manufacturers: E2V sensors in the center and ITL sensors on the
edges. Their locations are easily distinguishable as they exhibit different features.
Three main patterns stand out: an ITL ``blob'', amplifier offsets on the E2V sensors,
and a ring pattern at the edge of the focal plane. While a detailed investigation of
all three features is possible, this proceeding focuses primarily on the ITL blob,
which reveals information about both PSF model performance and, unexpectedly, the
contribution of the optics to our image quality budget. See the next section for
details.

\subsection{On the ITL ``blob'' or is my telescope in focus?}

This ITL blob in the $\delta T/T$ map was already observed with the commissioning
camera and described in the DP1 paper (Rubin Observatory Team 2026~\cite{DP1Paper}). The solution adopted then,
shown in Figure~16 of the DP1 paper, was to allow more flexibility in the spatial
variation of the PSF model across a CCD by using a 4th order polynomial interpolation
instead of the standard 2nd order. As seen in that figure, this removes the ITL blob,
but it does not explain the physics behind this pattern that appears only in ITL
sensors.

\begin{figure}
	\centering
  \includegraphics[scale=0.25]{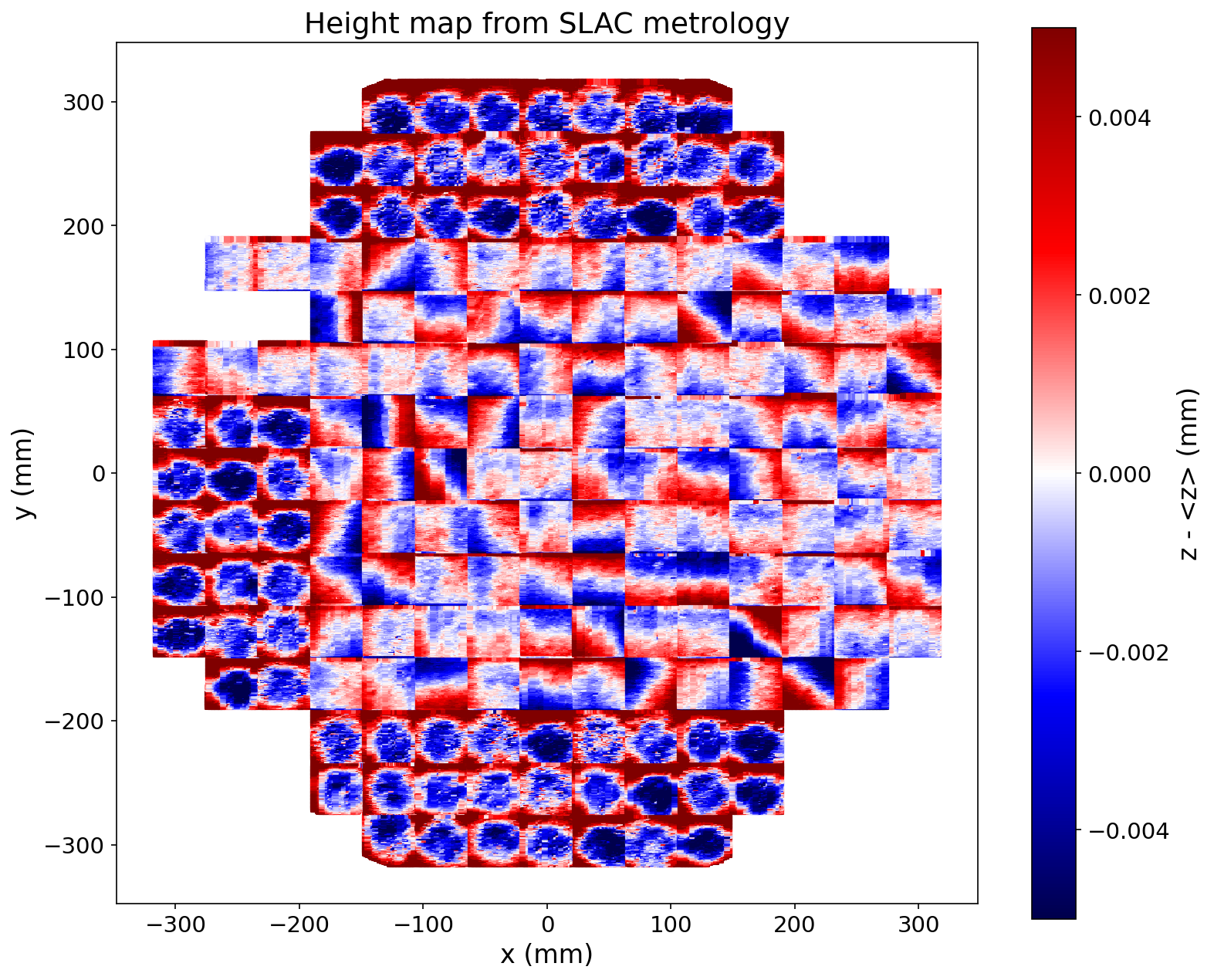}
  \includegraphics[scale=0.25]{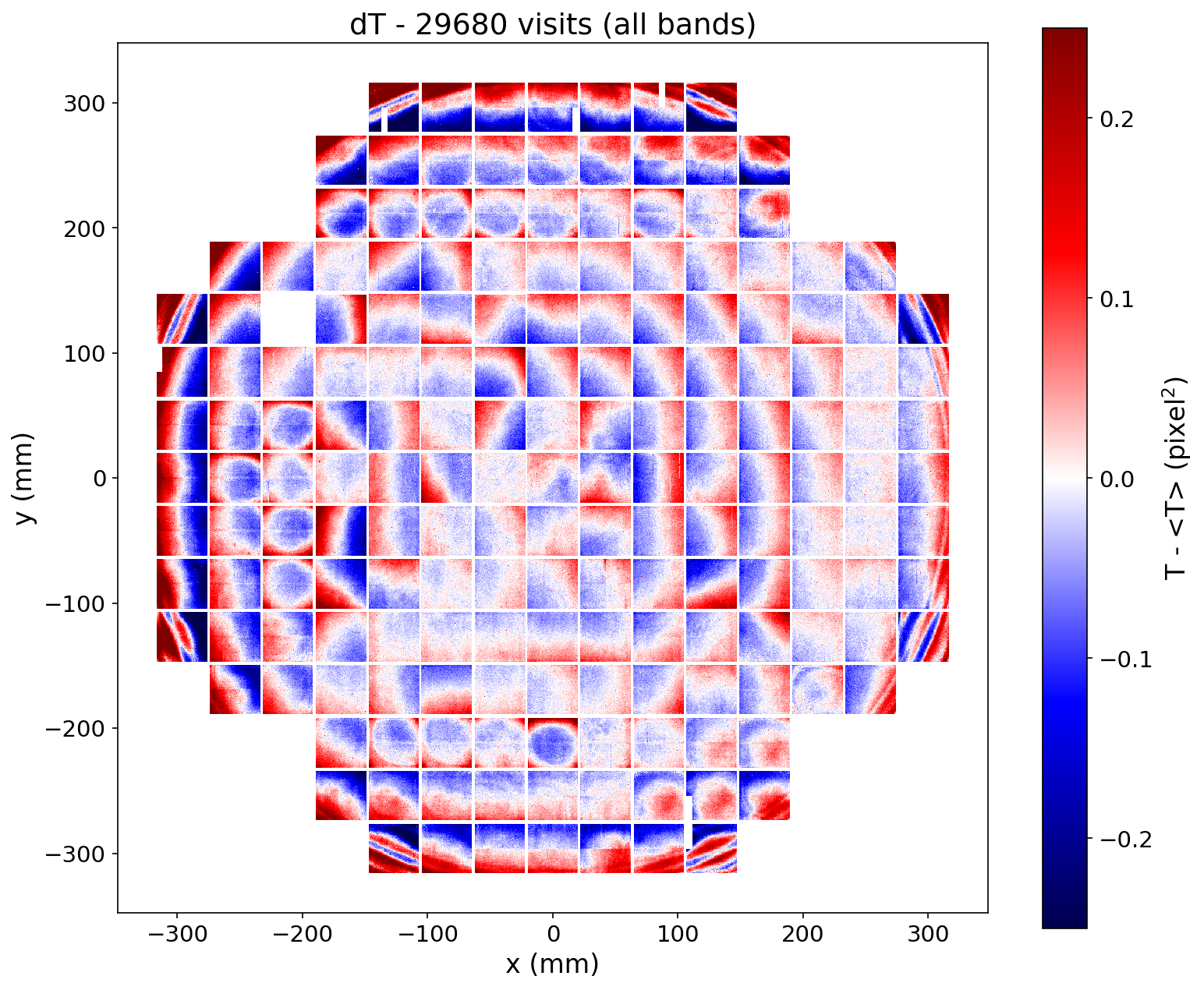}
  \vspace*{-4mm}
  \caption{Left: Height map of LSSTCam measured at SLAC. Right: PSF size residuals measured on thousands of science images before any
  interpolation is done (mean interpolation). Both Left and right match. 
}
  \label{HeightMap}
\end{figure}

To better understand the origin of this feature, instead of increasing to a 4th order
polynomial, we switch to a 0th order polynomial, as shown in Figure~\ref{HeightMap}. A
clear difference remains between E2V and ITL sensors: the ITL sensors exhibit a curved
circular pattern, while the E2V sensors show a tilt-like shape. If we interpret this
as local variations in focus across the camera due to height variations or non-flatness
of the focal plane, it should simply reflect the physical shape of the sensor array ---
and indeed, this is what we observe. Figure~\ref{HeightMap} shows the height map of the
camera as measured in the lab at SLAC National Accelerator Laboratory (SLAC) (Roodman et al. 2024~\cite{Roodman2024}). It is straightforward to see that $\delta T/T$
is highly correlated with the height map; therefore, the ITL blob is due to the complex
surface shape of the sensors themselves.

However, why is this happening at all? When the height measurements were made at SLAC,
this was not a concern, so why is the height map now imprinted on our PSF residuals?
The answer is that the telescope was mostly out of focus until mid-December 2025. The
Active Optics System (AOS) was slightly off-target: it was reaching the focus value
requested, but that target was not the true optimal focus. This was corrected in
December 2025, but by then most of the data destined for DP2 had already been acquired
slightly out of focus. This explains why the 4th order polynomial interpolation was
needed to remove the blob. Further analysis is required to determine whether this
higher-order interpolation is still necessary for the on-sky data currently being
gathered with the corrected focus settings. 

\subsection{PSF residuals in sky coordinates}

Switching to a 4th order polynomial interpolation helps remove most of the features
in the $\delta T/T$ maps, as demonstrated in the DP1 paper (Rubin Observatory Team
2026~\cite{DP1Paper}). However, most analyses during the first year of the Rubin Observatory
commissioning were performed on a limited subsample of visits. The Rubin Observatory Data Management
began processing DP2 data in January 2026, with PSF fitting completed during the same
period. This represented an order of magnitude increase in data volume, totaling
approximately 30,000 visits. With this larger dataset, we were able to perform more
detailed band-specific analyses. One notable finding was a global offset in
$\delta T/T$ that depended on the observing band.

\begin{figure}
    \centering
    \includegraphics[scale=0.4]{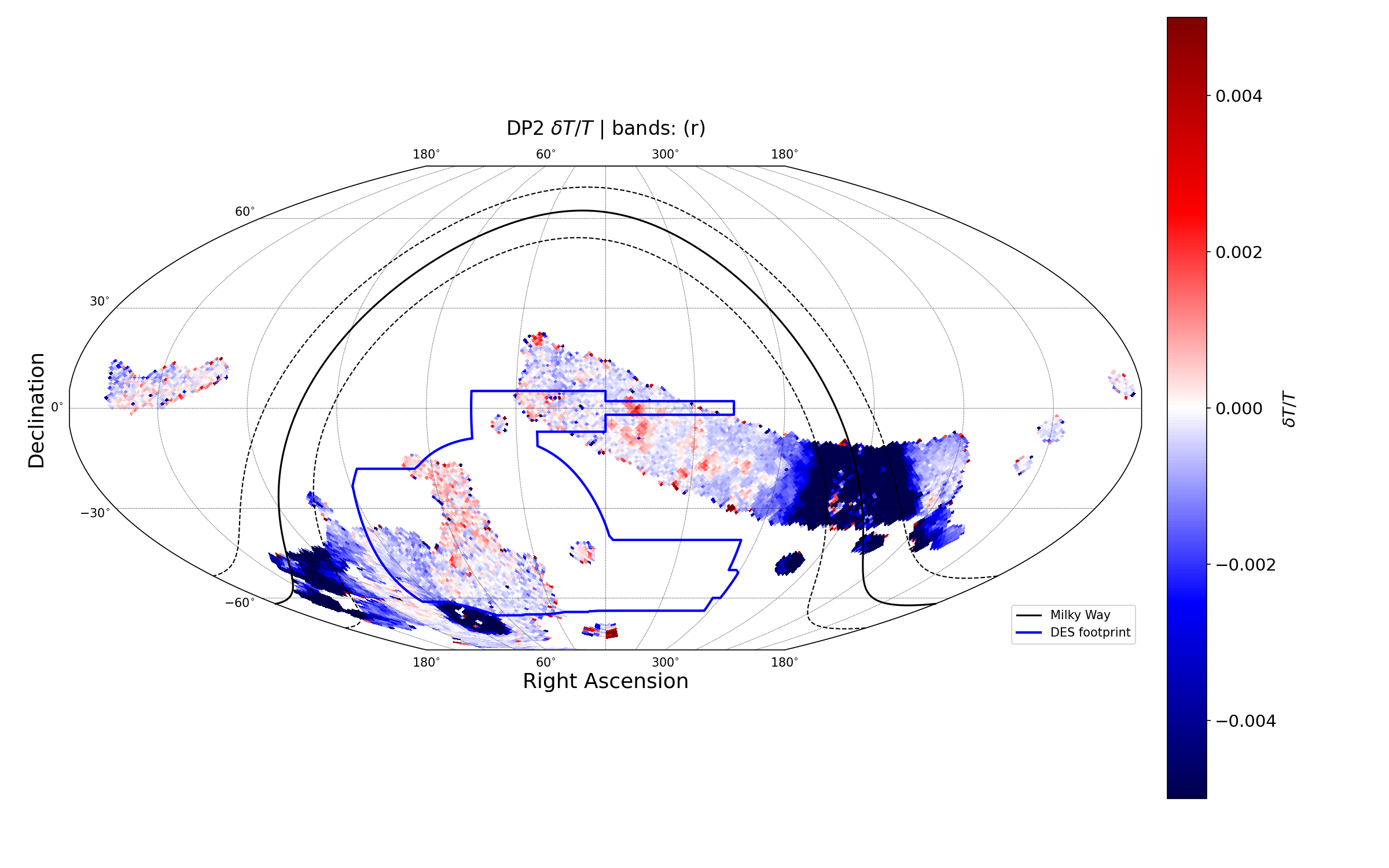}
    \vspace*{-4mm}
    \caption{$\delta T/T$ projected in sky coordinates for single visits in Data Preview~2. No image quality cut is applied.}
  \label{dTTSky}
\end{figure}

To understand the origin of this offset, we examine the spatial distribution of
$\delta T/T$ across the sky. Figure~\ref{dTTSky} shows a sky map of $\delta T/T$ along
with the positions of the Milky Way, the Large Magellanic Cloud (LMC), and the Small Magellanic Cloud (SMC). A clear correlation with
stellar density emerges: there is a larger bias in these high-density regions. The
offset therefore depends on how much time the survey spent observing high stellar
density fields in each band. We attribute this offset to problems in sky subtraction.
Areas with more stars are more challenging for sky estimation, and an incorrect
background estimate biases second moment measurements at low signal-to-noise and the
PSF model at high signal-to-noise. We have developed a proof of concept to address
this issue in PSF modeling, but it will not be available for Data Preview~2. It
should, however, be implemented for the first data release of LSST.

\subsection{Modeling the chromaticity of the PSF}

\begin{figure}
    \centering
    \includegraphics[scale=0.4]{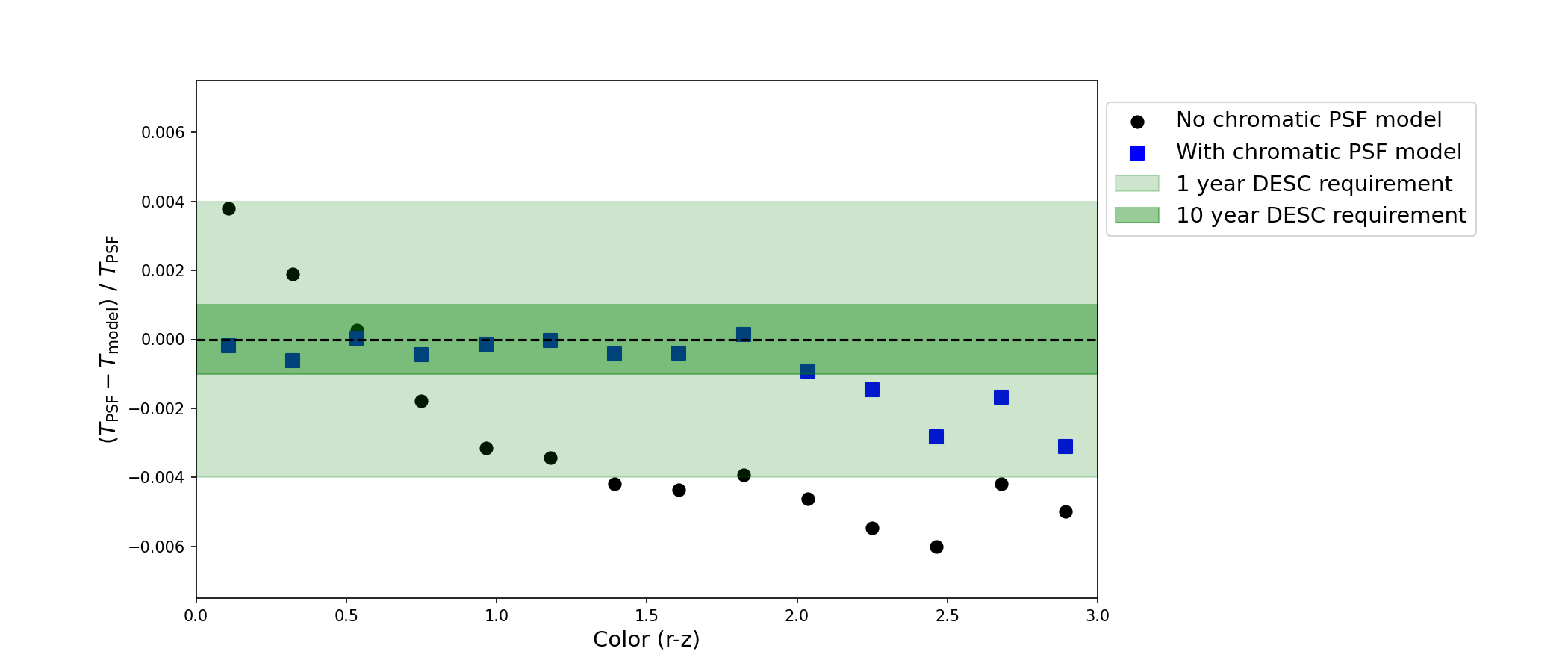}
    \vspace*{-4mm}
    \caption{$\delta T/T$ as a function of $r-z$ in Data Preview 1, with and without chromatic correction.}
  \label{ChromaticPSF}
\end{figure}

Differential chromatic refraction (DCR) and atmospheric turbulence introduce a
chromatic dependence of the PSF that manifests as a ``fatter-bluer'' effect. This
can be seen in Figure~\ref{ChromaticPSF} when not accounted for. Unlike the Hyper Suprime-Cam (HSC), LSSTCam does not
have an Atmospheric Dispersion Corrector, so this effect must be modeled explicitly.
In DES Y6 (Schutt et al. 2025~\cite{Schutt2025}), the approach was to add a third axis of interpolation
for the PSF model coefficients, where this third axis is stellar color. We
re-implemented this DES Y6 correction in the Rubin Observatory Data Management
pipeline. Figure~\ref{ChromaticPSF} shows the performance of this correction on Data Preview~1
data, which appears comparable to DES Y6. This may be sufficient for LSST Year~1, but
as suggested in Plazas et al. 2012~\cite{Plazas2012} and Meyers et al. 2015~\cite{Meyers2015}, it is likely
insufficient for LSST Year~10 and may require more sophisticated modeling.

\subsection{Improving astrometry for PSF}

\begin{figure}
	\centering
  \includegraphics[scale=0.16]{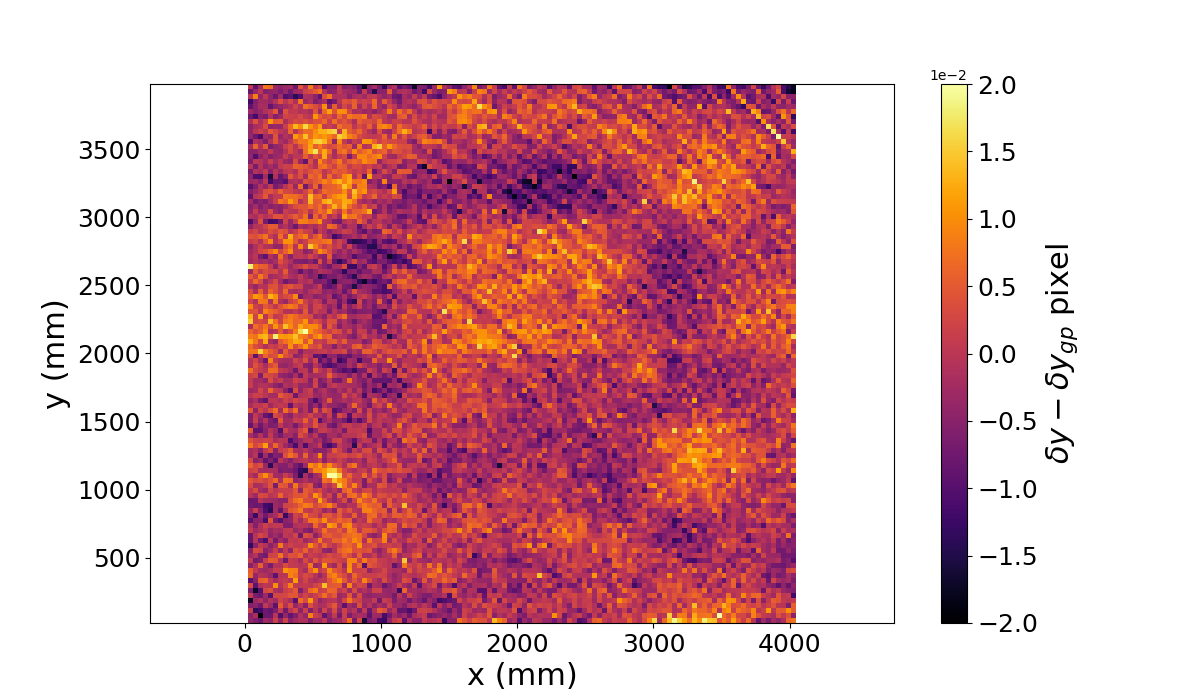}
  \includegraphics[scale=0.12]{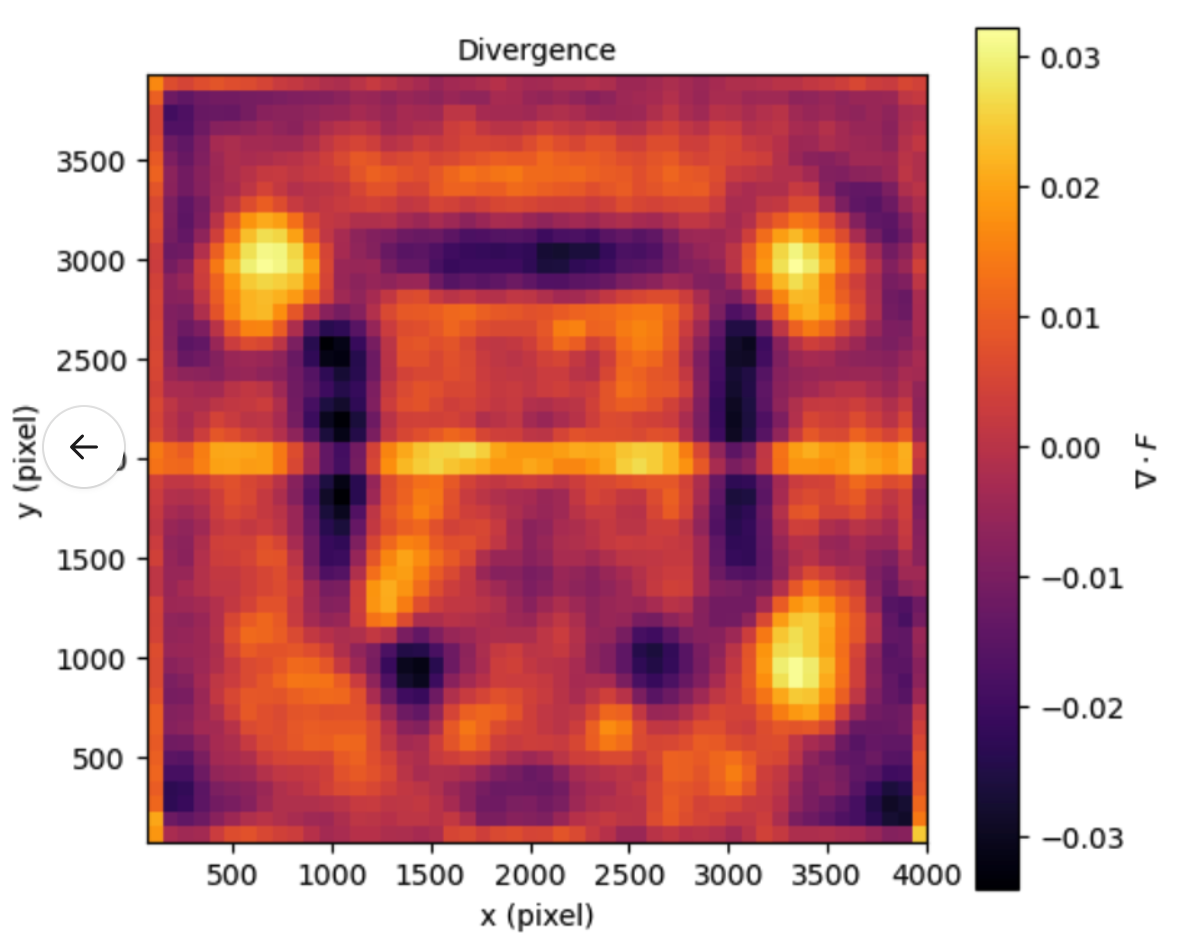}
  \includegraphics[scale=0.10]{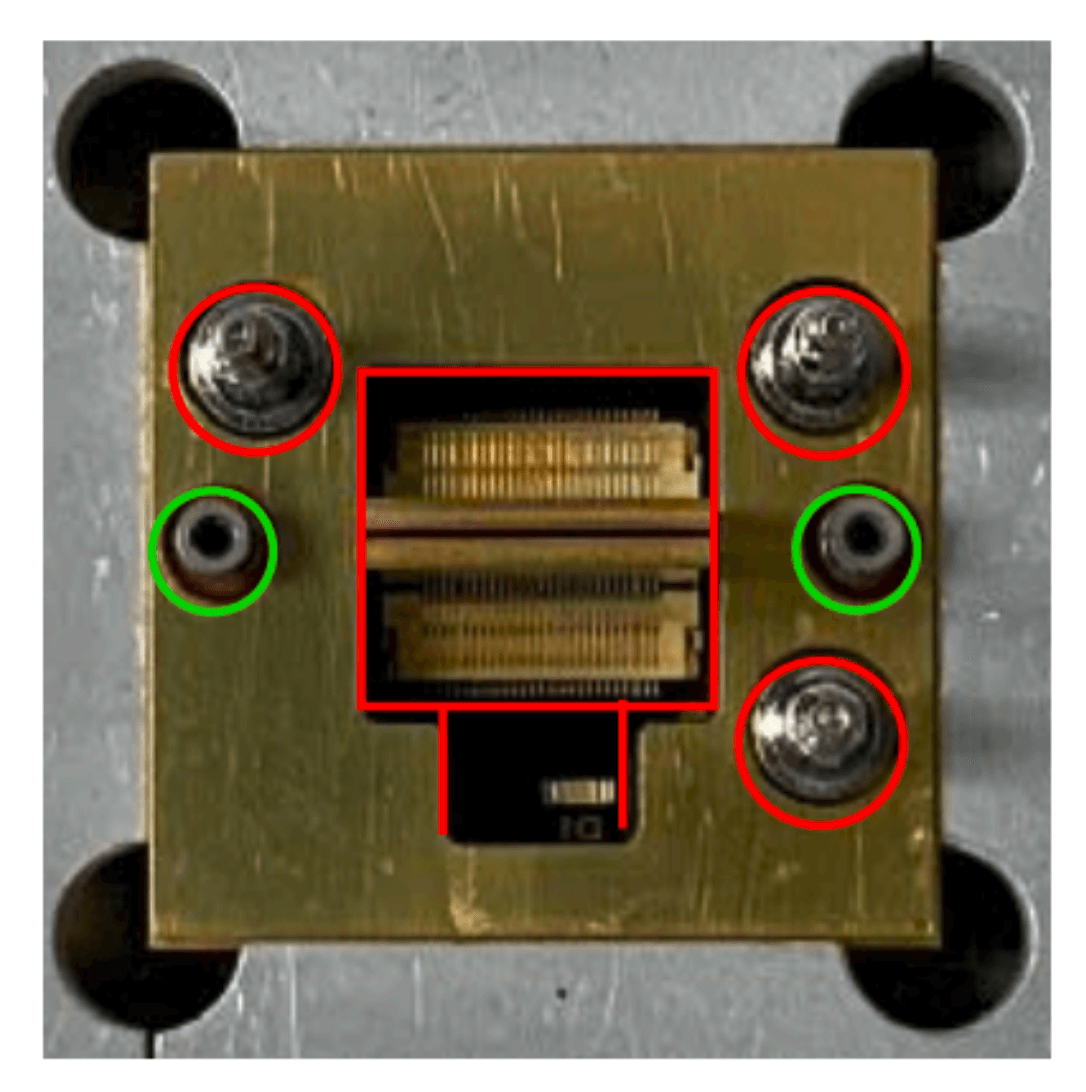}
  \vspace*{-4mm}
  \caption{
    Left: Stack of astrometric residuals in a single direction (y direction of the CCD).
    Two patterns are observed: the known and expected tree rings, but also a smoother pattern with slower variation. 
    Center: By averaging all the sensors together and by taking the divergence of the field,
    a clear structure appears that looks to be exactly the shape of the support of the detector (Right).
    This is for one of the 2 types of sensors that are on the focal plane (ITL), and the other type (E2V) exhibits
    similar structures. 
}
  \label{StaticDistortion}
\end{figure}

Another component of PSF modeling inspired by the Dark Energy Survey is working in
sky coordinates. One advantage of PIFF is that it can operate in sky coordinates,
thereby folding spatial variations due to CCD defects such as tree rings
(Plazas et al. 2014~\cite{Plazas2014}) into the World Coordinate System (WCS) transformation. This was one of the main
reasons for PIFF's success in the DES cosmic shear analysis. The astrometry in
the Rubin Observatory is modeled using the same approach as DES (Bernstein et al. 2017~\cite{Bernstein2017}). In
addition to characterizing CCD defects, as shown in Figure~\ref{StaticDistortion}, we implemented
a correction for atmospheric turbulence.

\begin{figure}
	\centering
	\includegraphics[scale=0.45]{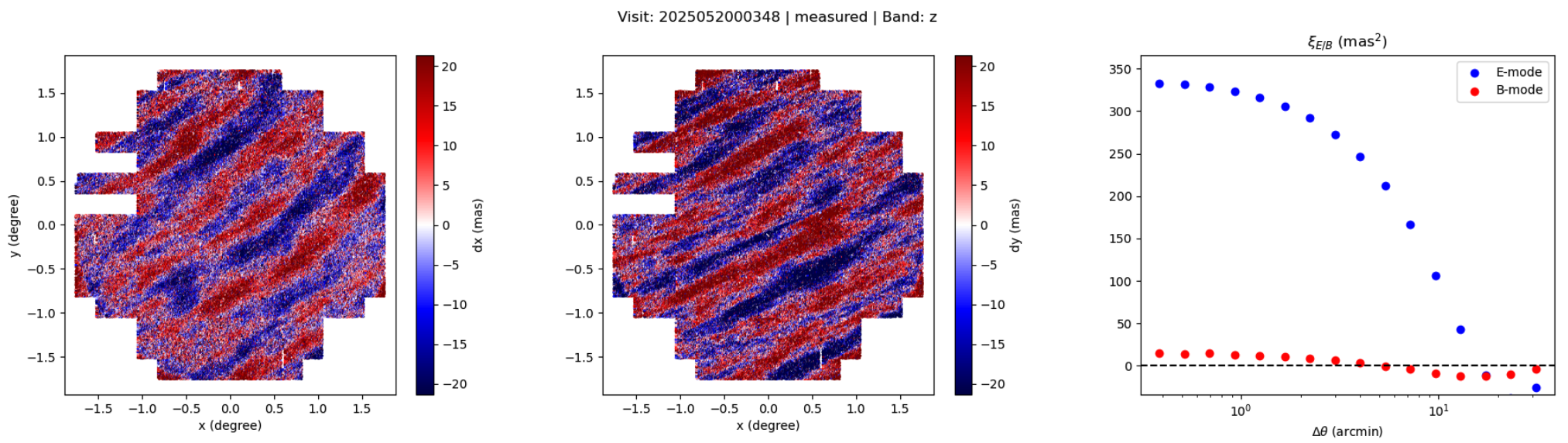}
	\vspace*{-2mm}
	\caption{
        Astrometric residuals from a single visit of LSSTCam. The two first panels on the left show each
        component of the astrometric residuals field and the last panel shows the E and B mode for this visit. The stripe patterns,
        which are pure E-mode, are characteristic of atmospheric turbulence and significantly higher than what was observed in previous surveys.
    }
	\label{visitAstro_LSST}
\end{figure}

\begin{figure}
	\centering
  \includegraphics[scale=0.47]{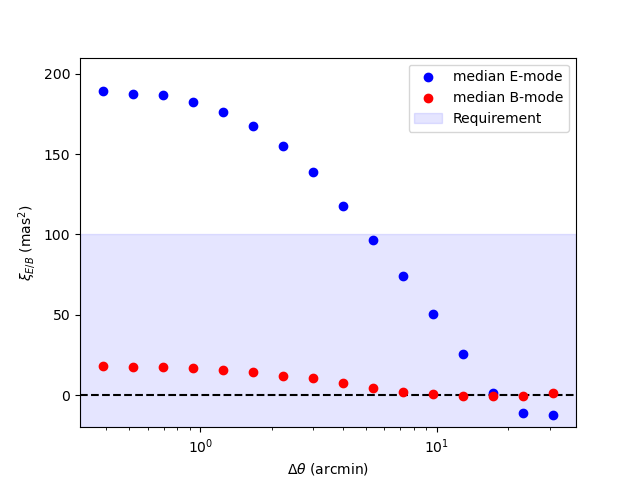}
  \includegraphics[scale=0.47]{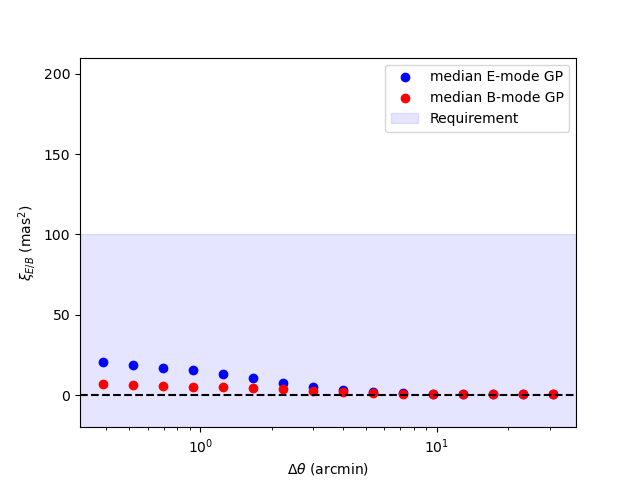}
  \vspace*{-4mm}
  \caption{Left: average across visits of E- and B- mode of the astrometric residuals
  field before GP modeling and on the right after GP modeling. GP removes more than 90\% of the observed
  variance due to the atmospheric turbulence. 
}
  \label{EB_GP_LSST_CORR}
\end{figure}

In DES and HSC, it was shown that astrometric residuals are dominated by atmospheric
turbulence (Fortino et al. 2021~\cite{Fortino2021}, Gomes et al. 2025~\cite{Gomes2025},
Léget et al. 2021~\cite{Leget2021}). A key difference with the Rubin Observatory
compared to DES and HSC is the 30-second exposure time, which is significantly shorter
than both surveys. The direct implication is that the variance is at least twice as
large in the Rubin Observatory as in DES due to the shorter integration time. However,
as demonstrated in DES and HSC, an anisotropic Gaussian Process (GP) can be used to correct
for this effect. We applied the same method as HSC (Léget et al. 2021~\cite{Leget2021}),
and as in HSC and DES, we are able to remove more than 90\% of the E-mode power
(see Figures~\ref{visitAstro_LSST} and~\ref{EB_GP_LSST_CORR}). This correction is part
of our full WCS solution but is not yet used in PSF modeling for Data Preview~2; it
will be incorporated in the first year data release.

\section*{Acknowledgments}

This material is based upon work supported in part by the National Science Foundation
through Cooperative Agreements AST-1258333 and AST-2241526 and Cooperative Support
Agreements AST-1202910 and 2211468 managed by the Association of Universities for 
Research in Astronomy (AURA), and the Department of Energy under Contract No. DE-AC02-76SF00515
with the SLAC National Accelerator Laboratory managed by Stanford University. Additional Rubin
Observatory funding comes from private donations, grants to universities, and in-kind support
from LSST-DA Institutional Members. This publication is based in part on proprietary Rubin Observatory
Legacy Survey of Space and Time (LSST) data, and was prepared in accordance with the Rubin Observatory
data rights and access policies. All authors of this publication meet the requirements for co-authorship
of proprietary LSST data. This research uses services or data provided by the Rubin Science Platform 
at NSF-DOE Vera C. Rubin Observatory, which is jointly funded by the U.S. National Science Foundation 
and the U.S. Department of Energy, Office of Science.

\section*{References}
\bibliography{main}


\end{document}